\documentclass[12pt]{iopart}
\usepackage{iopams}  
\usepackage{graphicx}
\begin{document}

\title[]{Energy gap revealed by low-temperature scanning-tunnelling spectroscopy of Si(111)-$7\times 7$ surface in illuminated slightly-doped crystals}

\author{A.B. Odobescu, S.V. Zaitsev-Zotov}

\address{Kotel'nikov IRE RAS, Mokhovaya 11, bld.7, 125009 Moscow, Russia}
\ead{arty@cplire.ru}
\begin{abstract}
Physical properties of Si(111)-$7\times 7$ surface of low-doped n- and p-type Si samples is studied in the liquid helium temperature region by the scanning-tunnelling microscopy and spectroscopy. Conduction required for the study is provided by illumination of the surface. Application of illumination removes completely the band bending near the surface and restores initial population of the surface states. Our results indicate the existence of the energy gap $2\Delta = 40\pm 10$~meV in intrinsically-populated Si(111)-$7\times 7$ surface.
\end{abstract}


\section{Introduction}
Si(111)-$7\times 7$ is one of the most extensively studied crystal surface of semiconductors. It has a relatively high surface density of states with the Fermi level sitting inside of one of the surface zones. The latter must lead to the metallic character of Si(111)-$7\times 7$ \cite{oura}. Experimental studies have demonstrated, however, contradictory behaviour. Some authors report metallic properties of Si(111)-$7\times 7$ \cite{losio}, whereas others observe the dielectric gap with the value varying between 70 meV \cite{modesti} and 1 eV \cite{myslivecek}. Non-metallic behaviour is also found in the transport measurements \cite{tanikawa}. Revealed contradictions result in significant increase of number of publications devoted to physical properties of Si(111)-$7\times 7$ surface. 

In general, low-temperature scanning-tunnelling microscopy (STM) and spectroscopy (STS) studies of Si(111)-$7\times 7$ surface are carried out with using heavy-doped Si (typical resistivity $\rho\sim 0.01 - 0.001$ $\Omega$cm). Such a high doping level is necessary to provide a reasonable conduction, especially in the low-temperature range. But the effect of doping on the surface concentration of current carriers is not negligible. As it will be argued below, a typical variation of the surface current carriers concentration is of the order of its value in heavy-doped crystals. Thus, in principle the effect of doping may be responsible for variation of physical properties of Si(111)-$7\times 7$ surface obtained  by different groups, so it devotes a special consideration. 

Here we  report the results of our STS study of energy spectrum of Si(111)-$7\times 7$ surface in slightly doped n- and p-type Si at temperatures 5~K and 78~K under illumination which was used to remove bands bending near the surface and provide STM measurements at 5~K. We also demonstrate the possibility of STM and STS measurements in nominally dielectric Si samples in a new mode when the external circuit contacts to the crystal surface. The results confirm the existence of the dielectric gap $2\Delta=40\pm 10$~meV in the low-temperature region on p-type and n-type crystals, in qualitative agreement with the earlier observation \cite{modesti}.

\section{Experimental}
We studied n- and p-type Si with $\rho = 1$ $\Omega$cm. Such crystals are insulating in the low-temperature region. Conduction required for STS and STM study can be achieved by illumination of the surface by light with the photon energy above the Si energy gap. Relatively lower doping level provides smaller band bending with respect to heavy-doped Si studied usually, and larger life-time for non-equilibrium carriers excited by a general  white light LED with the estimated light intensity $\sim 10^{-4}$ W/cm$^2$ at a sample position. The illumination can also be used for additional flattering of the band structure near the surface (see below). 

High-quality Si(111)-$7\times 7$ surfaces were prepared {\it in situ} with the standard procedure of 10-hours degassing at $T \approx 900$~$^o$C DC followed by flashing up to 1250$^o$C for 10 seconds and slow computer-controlled cooling-down.  All the experiments were performed in LT UHV Omicron STM at a base vacuum level below $10^{-10}$~Torr. We used cut platinum and electrochemically etched tungsten STM tips cleaned with focused electron beam in ultra-high vacuum conditions. We make measurements on several  p- and n-type silicon samples with different  Pt and W tips, to receive reproducible results. Relatively low conduction of studied surface does not allow to use fast voltage sweep, so we use 500 voltage point in each voltage sweep direction and 40 ms signal integration time for each voltage point. With such a setting no noticeable hysteresis between back and forth voltage sweep was observed. The I-V curves presented here were measured over the adatom in the center of a side of Si(111)-$7\times 7$ unit cell and averaged over series of measurements consisting of 10 individual cycles for survey curves and  50 for detailed curves. Similar results were obtained in the measurements performed over the corner adatom and the corner hole positions. Finally it took about half an hour to collect the data for a single detailed curve. The drifts at liquid helium temperature are insignificant and the STM tip is stable at one point during one measurement.  Differential conduction was calculated numerically from the resulting I-V curve.

\section{Results}
\begin{figure}
\centerline{\includegraphics[scale=.6]{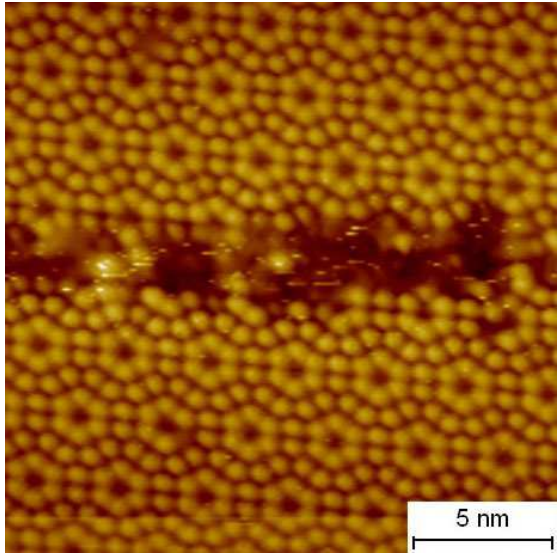}}
\caption{STM image of Si(111)$7\times 7$ surface (p-type, $\rho = 1$ $\Omega$cm, set points are $U_t = 2$~V and $I_t = 100$~pA, $T = 5$~K). The illumination is on in the bottom and top parts of the image and is off in the central part.}
\label{si_image}
\end{figure}

Fig.~\ref{si_image} shows an STM image of illuminated Si(111)-$7\times 7$ surface obtained at the helium temperature. The middle-part of the image was taken without illumination. It is clearly seen that switching the illumination off results in destruction of the reconstruction (image center) which was absent before this procedure. This indicates impossibility of STM study of low-doped Si samples without an additional conduction-improving action. Thus, illumination extends the range of doping level appropriate for STM study. 

\begin{figure}
\includegraphics[scale=.4]{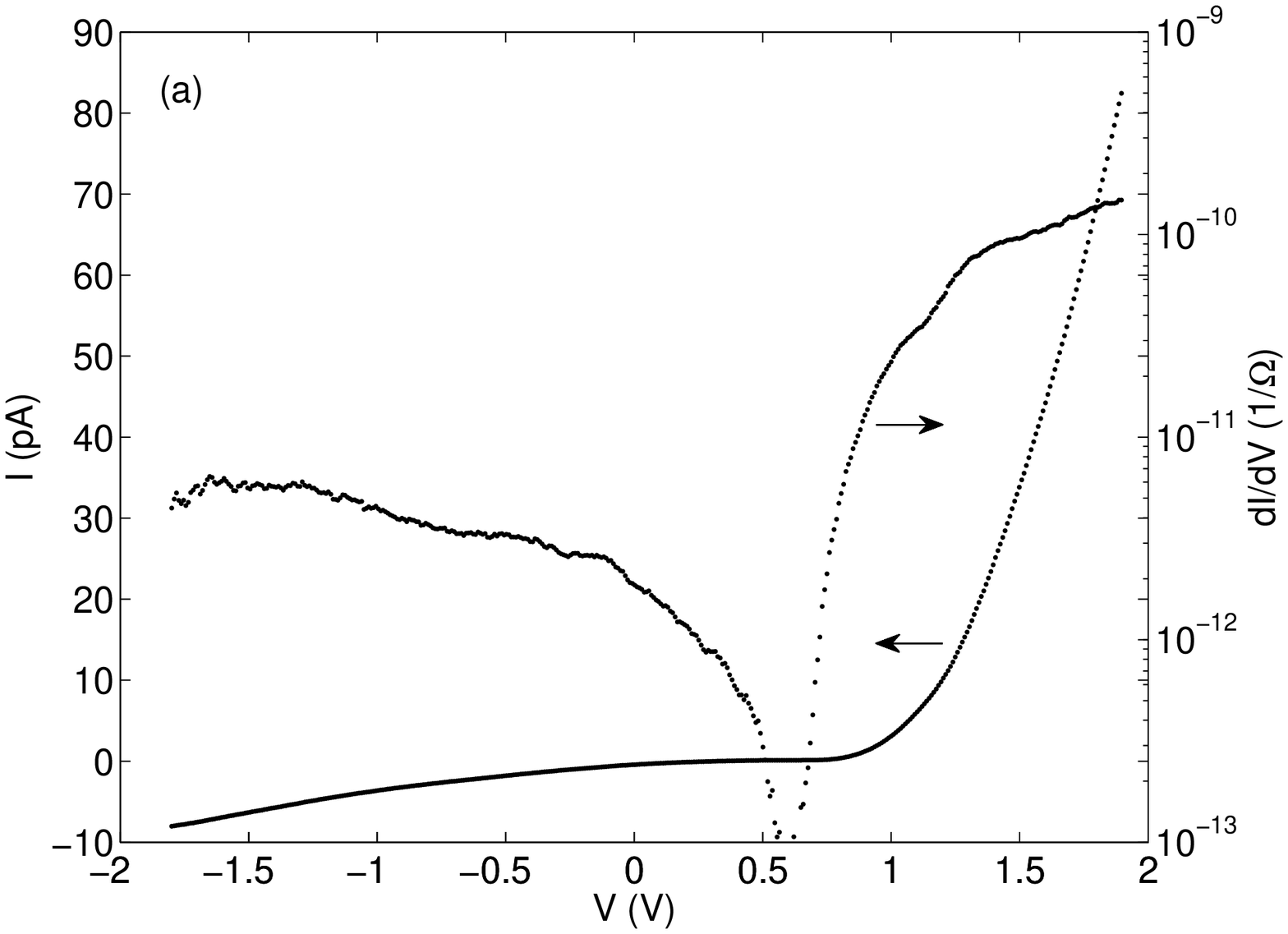}
\includegraphics[scale=.4]{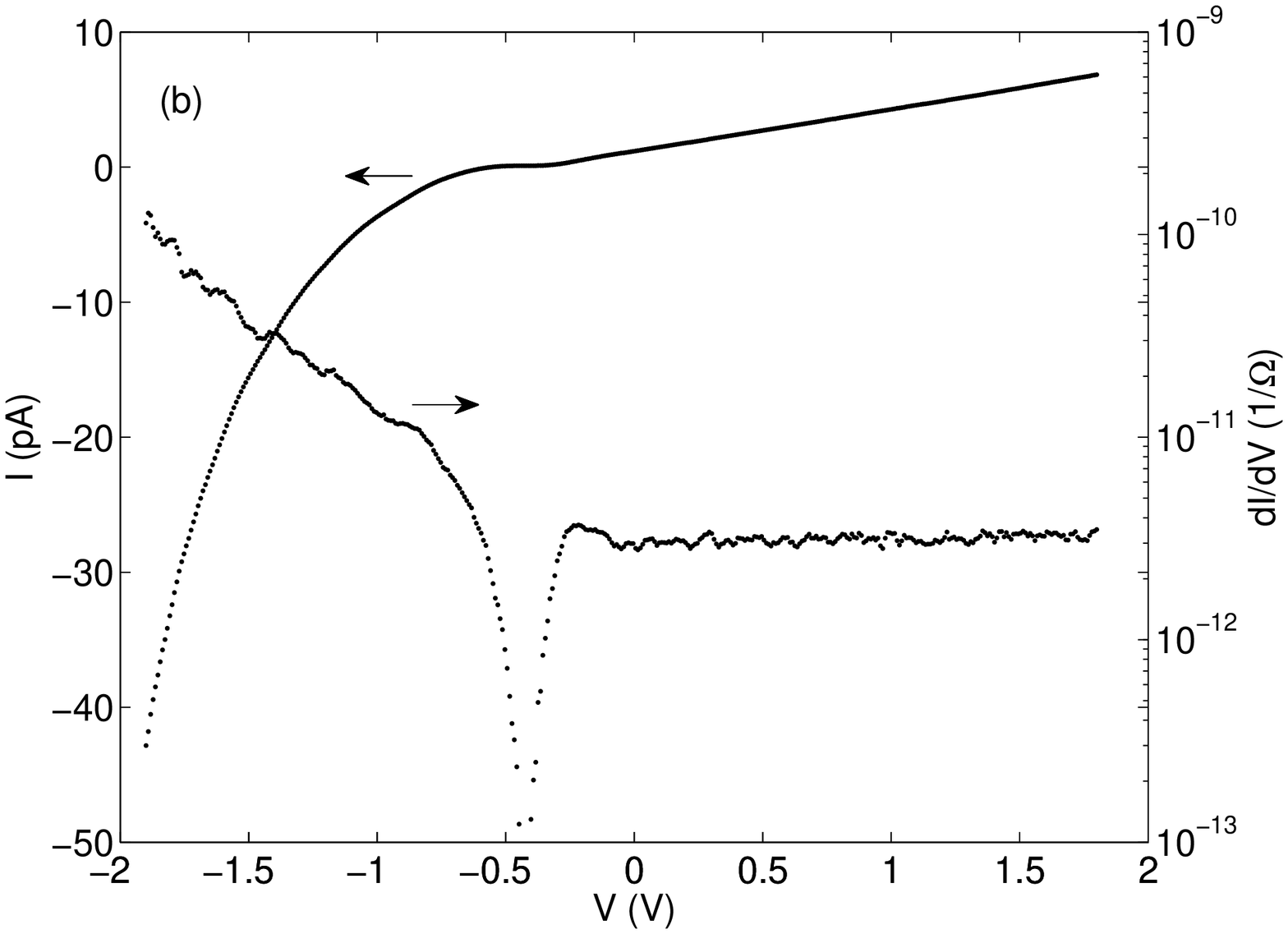}
\caption{I-V and $dI/dV$ curves of illuminated Si(111)-$7\times 7$ surface ((a) p-type, $\rho = 1$ $\Omega$cm; (b) n-type, $\rho = 1$ $\Omega$cm) at $T = 5$ K. The energy gap for both samples
shifted by photovoltage is clearly seen in $dI/dV$ curve. Positive voltage means the positive potential of the sample.}
\label{ivs}
\end{figure}

Typical low-temperature I-V and $dI/dV$ curves of illuminated Si(111)-$7\times 7$ surface of p- and n-doped crystals are shown in Fig.~\ref{ivs}. The curves indicate the presence of the energy gaps at non-zero voltage positions. Note that the features of Si(111)-$7\times 7$ surface (peaks S1, S2 {\it etc.}) are not seen on $dI/dV$ curves because of relatively low conduction of the sample \cite{myslivecek}. However the low value of conduction does not affect the gap position and its value determined from the data obtained at much smaller currents. 

\begin{figure}
\includegraphics[scale=.4]{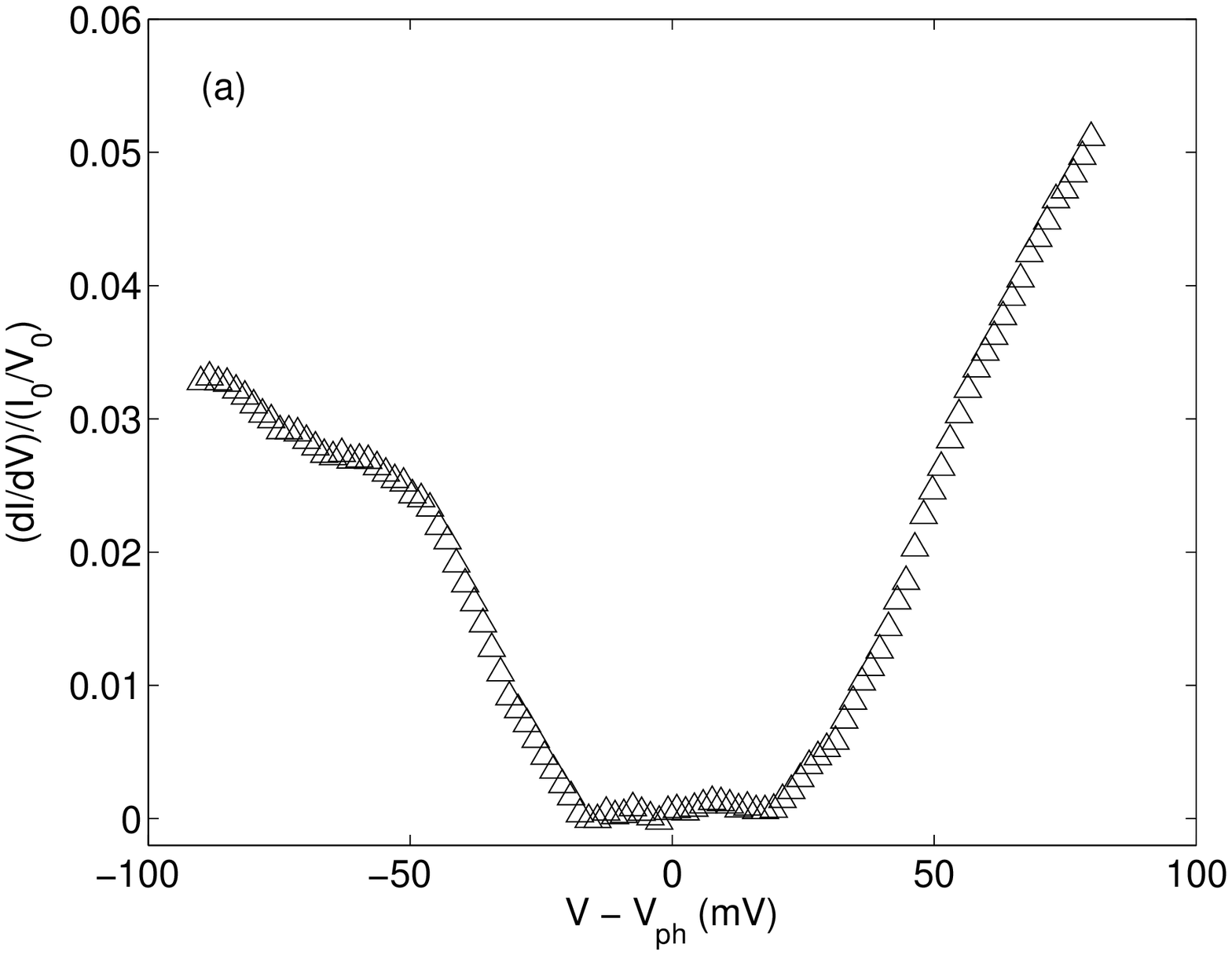}
\includegraphics[scale=.4]{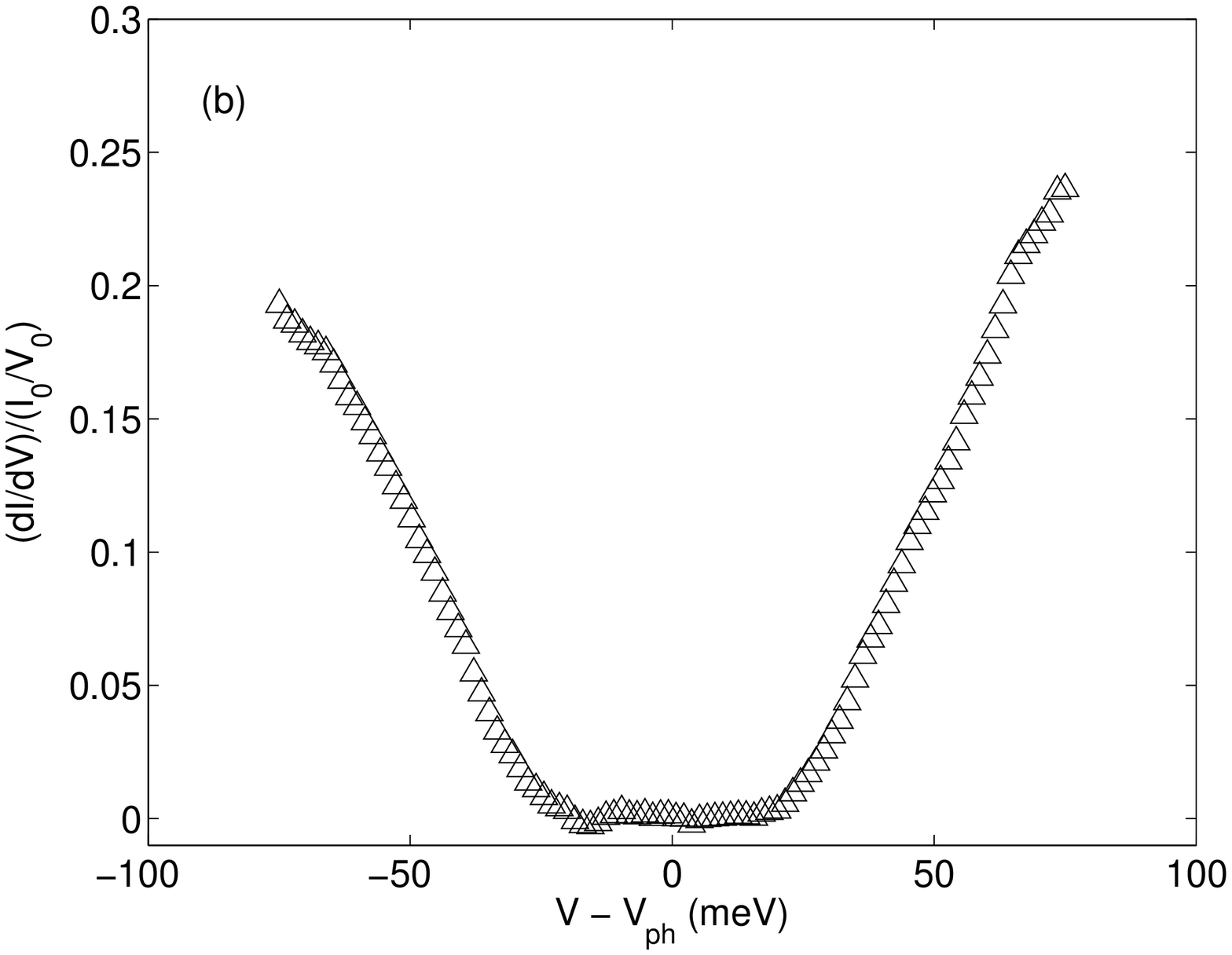}
\caption{$dI/dV$ curves of illuminated Si(111)-$7\times 7$ surface ((a) p-type, $\rho = 1$ $\Omega$cm; (b) n-type, $\rho = 1$ $\Omega$cm) in the gap region at $T = 5$ K. Set points are $V_0 = 2$~V and $I_0 = 100$~pA,}
\label{gaps}
\end{figure}

Fig.~\ref{gaps} provides more detailed view of the gap regions measured with 1 mV voltage resolution. The gap edges are relatively sharp, the gap values in both cases are the same, $2\Delta = 40 \pm 10$~meV. The results allow to determine the positions of the middles of the gaps: they are shifted from zero voltage by $+0.6$~V and $-0.4$~V for p-type and n-type respectively, due to the effect of surface photovoltage. The detailed explanation is given in the discussion section.  

\begin{figure}
\centerline{\includegraphics[scale=.4]{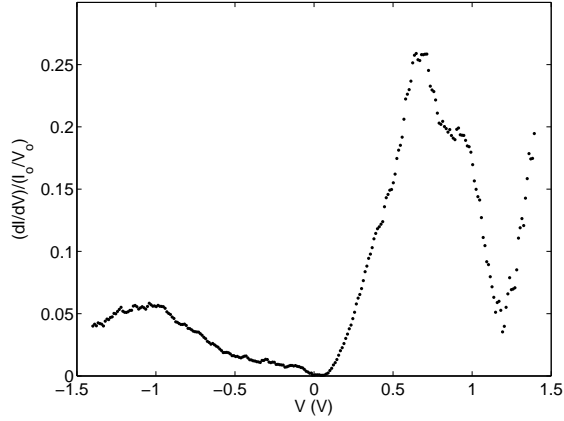}}
\caption{$dI/dV$ curve of clean Si(111)-$7\times 7$ surface surrounded by deposited In contact film, taken under illumination (p-type sample, $\rho  = 1$ $\Omega$cm) at $T = 5$~K. The energy gap exists at zero voltage $2\Delta =  50\pm 10$~meV, no photo-voltage shift is observed.}
\label{ptypegaponoff}
\end{figure}

In addition, we studied the energy spectrum of Si(111)-$7\times 7$ sample with the surface partially covered by In. The procedure of In deposition was the following. The STM tip was placed 10~nm from the surface, thus creating a shadow mask, the STM feedback circuit was switched off. Then a small amount of In was deposited on the surface at 5~K, the STM feedback circuit was switched on and STM return into a usual regime. The In film thickness obtained from the STM profile data was $\sim$ 10 nm. As a result, a fragment of clean Si(111)-$7\times 7$ surface surrounded by In layer was prepared. We found that such a procedure provides enough surface conduction for STM/STS study of the clean fragment of the surface even without illumination. Illumination, however, enables further conduction increase and facilitates observation of the energy gap. Such a coverage provides an electrical contact to  Si(111)-$7\times 7$ surface and eliminates the effect of surface photo-voltage. The results are presented at Fig.~\ref{ptypegaponoff}.

\begin{figure}
\centerline{\includegraphics[scale=.4]{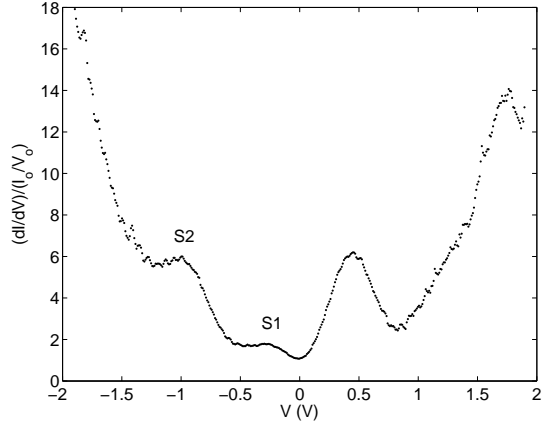}}
\caption{Normalized $dI/dV$ curves of Si(111)-$7\times 7$ structure without illumination (p-type sample, $\rho = 1$ $\Omega$cm) at $T = 78$~K. No energy gap is observed. S1 and S2 denote the surface bands in accordance with the general notation \protect\cite{oura}.}
\label{ltgap}
\end{figure}

At $T= 78$~K the effect of illumination on I-V curves is many orders smaller and does not affect the shape of I-V curves. At this temperature the energy gap disappears and the Fermi level sits inside the surface band providing thereby the metallic conduction of the surface (Fig.~\ref{ltgap}), in agreement with the general results \cite{oura}. 

\section{Discussion}
\begin{figure}
\centerline{\includegraphics[scale=.64]{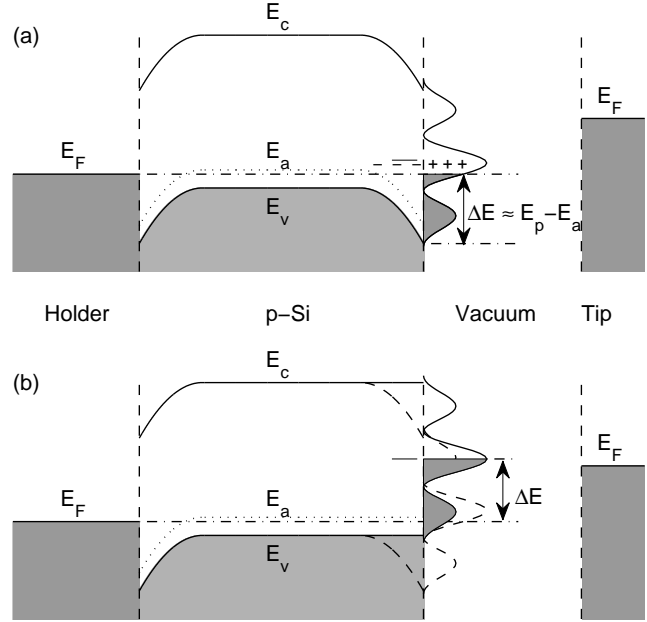}}
\caption{Energy diagrams of the studied system: a) metal holder - p-type Si - Si(111)-$7\times 7$ surface without illumination and STM tip (Schottky barrier and the respective band bending are present). Signs + and $-$ denote the space charge, the surface states are partially depleted; b) Effect of illumination on p-type Si - Si(111)-$7\times 7$ surface. Illumination suppresses the Schottky barrier and flatten the bands. Here $E_c$, $E_v$ are  conduction and valence bands edges, $E_a$ is impurity level, $E_p$ is pinning level for Si(111)-$7\times 7$ structure, $V_{sb}$ is the tunnelling voltage, $\Delta E$ is the band bending, $E_f$ denotes the Fermi level.}
\label{en_dia}
\end{figure}

The Fermi level is strongly pinned for both p- and n-type Si(111)-$7\times 7$ surface due to a substantial density of states near Fermi level \cite{himspel}. The pinning position lies near the middle of the bulk band gap at 0.65 eV above valence band maximum \cite{himspel}. Pinning of the Fermi level is provided by a variation of surface electron concentration required to screen the electric field produced by charged impurities in the space charge region, shown schematically in Fig. \ref{en_dia}. The variation of the surface current carrier concentration can be estimated as the total charge density of the space charge region per unit surface area  
$$
\Delta Q \approx \textit{d}\cdot N_{a}
$$
(for p-type silicon), where $N_{a}$ is the bulk concentration for acceptor atoms, depends on doping level, $\textit{d}$ is the depletion layer thickness. In its turn $d$ can be estimated as
$$
\textit{d}\approx\sqrt{\frac{2\epsilon\epsilon_{0}\Delta E}{\textit{e}^{2}\textit{N}_{a}}},
$$ 
where $\Delta E=E_{p}-E_{a}$, $E_p$ is the pinning position of Fermi energy on the surface, $E_p=0.65\pm0.05$ eV for Si(111)-$7\times 7$ \cite{himspel}, $ E_{a}$ is the bulk Fermi level which is  approximately the bulk acceptor energy level 0.045 eV (both $E_p$ and $E_a$ are shown here with respect to the top of the valence band), $\epsilon =11.7$ is the relative permittivity of Si. The Si(111)-$7\times 7$ surface has 5 electrons per unit cell \cite{ortega}, thus the surface current carrier concentration can be estimated as $ N_s \approx 8 \cdot 10^{13}$  $cm^{-2}$. For heavy-doped silicon samples widely used in low-temperature STM/STS studies, with typical resistivity $\rho\sim 0.01 - 0.001$ $\Omega$cm, the impurity concentration $N_a$ is an order of $10^{19} \div  10^{20}$ cm$^{-3}$ \cite{Sichapter}. The respective concentration change can be estimated as $\Delta N_s \approx 0.9\cdot 10^{13} \div 3\cdot 10^{13}$ cm$^{-3}$, which amounts $10\%$  to $40\%$ of the nominal value. For slightly doped silicon with $\rho \gtrsim 1$~$\Omega$cm  this change is $\Delta N_s \approx 8 \cdot 10^{10}$~$cm^{-2}$  smaller than 1\%. 

Thus, the effect of variation of surface carrier concentration in heavily doped crystals may be strong and should be taken into account. As it will be showed below, external illumination  removes the effect of surface carrier concentration change and allows to investigate the intrinsic properties of  Si(111)-$7\times 7$ surface.

Illumination-induced changes of the energy structure near the surface are shown schematically in Fig. \ref{en_dia}. In the dark, pinning of the Fermi level at the surface leads to formation of the Schottky barrier with the respective band bending $\Delta E$. Non-equilibrium carriers produced by illumination are separated by the electric field near the surface reducing thereby this field. At sufficiently large illumination intensity the field is almost completely suppressed and the bands are practically flat. The surface potential is shifted by $\Delta E$, the band bending without illumination (i.e. without the Schottky barrier) and restore initial current carrier concentration.

The surface photovoltage shifts the whole I-V curve by $V_{ph} \leq \Delta E/e$, where $e$ is the elementary charge,  in positive direction $\Delta E = E_p - E_a$, $\Delta E \approx 0.6$~eV for p-type samples and in negative direction $\Delta E = E_g - E_p - E_d$, $\Delta E \approx 0.47$~eV for n-type samples, where $E_p$ is the Fermi surface pinning level for $7\times 7$ structure (0.65~eV) \cite{himspel}, $E_a$ and $E_d$ are acceptors and donors energy levels for B and P dopants, $E_g$ is the silicon bulk gap 1.17~eV at $T=5$~K. The observed values of the photo-voltage shifts ($+0.6$~eV for p-type and $-0.4$~eV for n-type samples) practically coincide with the estimated ones $+0.6$~eV and $-0.47$~eV. That means that illumination almost completely suppresses the Schottky barrier.

Odd number of electrons per unit cell implies metallic conduction of Si(111)-$7\times 7$ surface. Observation of the energy gap at the Fermi level testifies importance of electron-electron correlations, as it was already discussed in numerous papers (see {\it e.g.} \cite{losio,modesti,ortega} and references therein). Possible scenarios of gap opening include Mott-Hubbard mechanism assisted by electron-phonon interaction \cite{modesti}, formation of two-dimensional Wigner crystal or density waves. Note, that the energy gap value of Si(111)-$7\times 7$ surface is the same in slightly doped p- and n-type samples, but is almost 2 times smaller than the gap value reported in Ref.~\cite{modesti} for heavily-doped samples. In most scenarios the smaller gap value corresponds to a bigger bare electron density of states of Si(111)-$7\times 7$ surface at the Fermi level. An additional contribution of random potential caused by doping impurities into the gap value cannot be excluded. Further study is necessary to distinguish different possibilities.

\section{Conclusion}
In conclusion, we study the low-temperature energy structure of Si(111)-$7\times 7$ surface. We argue that physical properties of the surface should depend essentially on the doping level, therefore we performed our measurements on low-doped Si samples. Illumination allows to extend the range of doping level appropriate for STM study, removes almost completely the band bending near the surface and reveals intrinsic properties of studied surface. Our results indicate the existence of  the energy gap  $2\Delta = 40 \pm 10$~meV in Si(111)-$7\times 7$ surface at $T = 5$~K for both p- and n-type low doped silicon crystals.

The work was supported by the Russian Foundation for Basic Research, the program of Presidium of RAS and by Russian Ministry for Science and Education (contract \# 16.513.11.3066). These researches were performed in the frame of the CNRS - RAS RFBR Associated European Laboratory Physical properties of coherent electronic states in condensed matter between Institut Neel, CNRS and Kotel`nikov IRE RAS.

\end{document}